\documentclass[preprint,aps,nofootinbib]{revtex4}
\usepackage{graphicx}
\usepackage{amsmath}
\usepackage{amsfonts}
\usepackage{amssymb}
\usepackage{color}%
\providecommand{\U}[1]{\protect\rule{.1in}{.1in}}

\newcommand{\f}{\begin{equation}}
\newcommand{\ff}{\end{equation}}
\newcommand{\fa}{\begin{eqnarray}}
\newcommand{\ffa}{\end{eqnarray}}

\begin{document}
\title{Thermal non-Gaussianitity in holographic cosmology }
\author{Yi Ling}
\email{yling@ncu.edu.cn}
\author{Jian-Pin Wu}
\email{jianpinwu@yahoo.com.cn} \affiliation{ Center for Relativistic
Astrophysics and High Energy Physics, Department of Physics,
Nanchang University, 330031, China}

\begin{abstract}

Recently it has been shown that the thermal holographic fluctuations
can give rise to an almost scale invariant spectrum of metric
perturbations since in this scenario the energy is proportional to
the area of the boundary rather than the volume. Here we calculate
the non-Gaussianity of the spectrum of cosmological fluctuations in
holographic phase, which can imprint on the radiation dominated
universe by an abrupt transition.  We find that if the matter is
phantom-like, the non-Gaussianity $f_{NL}^{equil}$ can reach ${\cal
O}(1)$ or even be larger than ${\cal O}(1)$. Especially in the limit
$\omega\rightarrow -5/3$, the non-Gaussianity is very large and
negative. Furthermore, since the energy is proportional to the area,
the thermal holographic non-Gaussianity depends linearly on $k$ if
we neglect the variation in $T$ during the transition (fixed
temperature).

\end{abstract}
\maketitle

\section{Introduction}

Inflation is an extremely successful model in solving the horizon
and flatness problems in the standard hot big bang cosmology
\cite{accelerate}. Furthermore inflation shows us that the CMB
anisotropy \cite{CMBobserve,WMAP,WMAPcon2} and the observed
structures in the universe
\cite{Mukhanov81,Guth82,Hawking82,Starobinsky82,Bardeen83} are
results of the quantum fluctuations occurring in the very early
stage of the universe. Its prediction of a nearly scale invariant
spectrum has been confirmed very well in the experiments
\cite{WMAP5}. Nevertheless, the possibility that primordial {\it
thermal} fluctuations might seed the structure of our universe is
also an intriguing alternative to quantum fluctuations
\cite{peeb,steph,rob,mag-pog}. Unfortunately the spectral index of
thermal fluctuations is either too red ($n_{s}=0$) or too blue
($n_{s}=4$), failing to generate a nearly scale invariant spectrum
\cite{thermalmilne, thermalloop, thermalholography1}. Therefore,
thermal scenarios call for new physics in order to produce nearly
scale invariant fluctuations. For example, a thermal scenario in
which the effect of new physics to change the equation of state of
thermal matter, can produce nearly scale invariant spectrum. This
happens in non-commutative inflation
\cite{steph,non-commutative1,non-commutative2}, and also in LQC
\cite{ddrlqc,thermalloop}. Another happens in the Hagedorn phase for
certain types of string gas \cite{Hagedorn} or in holographic phase
\cite{thermalholography1}, in which the energy becomes strongly
non-extensive, specifically proportional to the area. In addition,
postulating a mildly sub-extensive contribution to the energy
density in Near-Milne universe \cite{thermalmilne}, the
scale-invariance spectrum can also be obtained. In this paper, we
focus on holographic thermal fluctuations and investigate the
holographic thermal non-Gaussianity.

This scale invariance suggests that the primordial perturbation
arising from a thermal holographic phase may be able to seed the
structure of observable universe, and lead to the anisotropy
observed in CMB as well. Thus it is interesting and significant to
ask what is the distinct feature of this holographic primordial
perturbations and which is testable in coming observations. There
are at least two possible observables, accessible in the near
future, that have the potential to rule out or support large classes
of models: non-Gaussianity and primordial gravitational waves. In
Ref. \cite{gravitywave}, Y. S. Piao has investigated the tensor
perturbation. He finds that the tensor perturbation amplitude has a
moderate ratio, which may be tested in coming observations. For
comparison, we can also see Ref. \cite{gravitywavesgc} for tensor
perturbation from string gas cosmology. Here we focus on the
non-Gaussianity in holographic cosmology.

In the WMAP convention, the degree of non-Gaussianity is
parameterized by $f_{\rm NL}$, which can be written as
\begin{equation}
\zeta=\zeta_g+\frac{3}{5}f_{\rm NL}\left({\zeta}_g^2-\langle
{\zeta}_g^2\rangle\right) ~,
\end{equation}
where $\zeta$ is the primordial curvature perturbation and $\zeta_g$
is its linear Gaussian part. The introduced factor $3/5$ is
convenient for cosmic microwave background comparisons
\cite{WMAPcon1,WMAPcon2}. However, this ansatz only describes the
most generic form of non-Gaussianity which is local in real space.
Theoretically, the non-Gaussianity estimator $f_{\rm NL}$ should be
defined by the three-point function $\langle
\zeta_{k_{1}}\zeta_{k_{2}} \zeta_{k_{3}} \rangle$ and has different
shapes. We usually focus on two cases. One is the local, squeezed
limit $k_{1} \ll k_{2}\simeq k_{3}$, denoted by $f_{\rm
NL}^{local}$. Another is the non-local, equilateral limit $k_{1}
\simeq k_{2} \simeq k_{3}$, labled by $f_{\rm NL}^{equil}$
\cite{shape}(for brief, we can also refer to \cite{shapeli}).

Recently, the WMAP 5-year data shows that at 95\% confidence level,
the primordial non-Gaussianity parameters for the local and
equilateral models are in the region $-9<f_{\rm NL}^{local}<111$ and
$-151<f_{\rm NL}^{equil}<253$, respectively \cite{WMAP5}. If this
result is confirmed by the future experiments, such as the Planck
satellite, then it will be a great challenge to many inflation
models. For example, in the simplest single field slow roll
inflation model, it has been found that $\mid f_{NL} \mid <1$
\cite{NG1,NG2}. Such a non-Gaussianity is too small to detect in the
near future. However, there are by now a number of different ways of
generating density perturbations from sources other than the
fluctuations of slow roll inflation, and in these models it is
possible to get larger and observable non-Gaussianity
\cite{DBI,K,Ghost,add}.

In Ref. \cite{thermalChen}, B. Chen {\it et al.} find that thermal
fluctuations can potentially produce large non-Gaussianity. They
develop a method to calculate the non-Gaussianity of the
fluctuations with thermal origin and apply it to study the
non-Gaussianity in string gas cosmology \cite{thermalSGC}. In this
paper, we intend to investigate the thermal non-Gaussianity in the
holographic cosmology.

The outline of this paper is the following. We first present a
brief review on the holographic thermal fluctuations in section
\textrm{II}. Then we use thermal correlation functions to
calculate the power spectrum and the non-Gaussianity in section
\textrm{III}. Finally, we discuss the conditions which may lead to
a large non-Gaussianity in this scenario, in comparison with those
in string gas cosmology in section \textrm{IV}.

\section{Brief review on the holographic cosmology}

A holographic version of thermal cosmology was originally proposed
by J. Magueijo {\it et al.} (for more details, see Ref.
\cite{thermalholography1}). They postulate that the very early
universe underwent a phase transition from a high temperature,
holographic phase (Phase \textrm{I}) to a usual low temperature
phase of standard cosmology (Phase \textrm{II}). In holographic
phase, the universe is disordered such that there is no classical
metric \cite{quantumspacetime}, but its thermodynamics may be
described by making use of the holographic principle
\cite{holography}. The spacetime geometry emerged only during the
phase transition such that in Phase \textrm{II} the universe may be
described in terms of Einstein gravity theory. The basic idea in
holographic cosmology is that owing to the existence of an early
holographic phase, in which a specific heat scales as the area of
the boundary, the thermal fluctuations which arise in holographic
phase can finally imprint a scale invariant spectrum on Phase
\textrm{II} through the phase transition \cite{thermalholography1}.

More explicitly, in holographic phase it is conjectured that
\begin{equation}\label{energyE}
  \langle E \rangle = bM_{pl}^{2}T\langle A
  \rangle~,
\end{equation}
where $b$ is some dimensionless constant, $M_{pl}$ is the Planck
mass, $T$ is the temperature, and $\langle A \rangle $ is the
expectation value of the area of the boundary. Thus, at fixed area,
using the relation $c_{A}=(\frac{\partial \langle E
\rangle}{\partial T})_{\langle A \rangle}$, the specific heat can be
expressed as
\begin{equation}\label{heat}
  c_{A}=b\frac{\langle A \rangle}{\hbar G}~,
\end{equation}
where $\hbar$ is the Plank constant and $G$ is the Newton constant.
Therefore, the specific heat at fixed area is proportional to the
area which leads to scale-invariant spectrum.

When the temperature falls to a critical temperature $T_{_{c}}$, the
phase transition begins and classical metric emerges. As the concept
of length is created, the area and volume can be expressed as
$\langle A\rangle = A = 4\pi R^{2}$, and $V(R) = \frac{4}{3}\pi
R^{2}$ as usual, where $R$ is the thermal correlation length. For
more precisely depicting the phase transition process, the
dependence of $R$ on $T$ was introduced \cite{thermalholography1},
\begin{equation}\label{RandT}
  \frac{R(T)}{l_{0}}=(\frac{T_{c}}{T_{c}-T})^{\gamma}~,
\end{equation}
where the relation is valid for $T\leq T_{c}$, $l_{0}$ is the
smallest scale but not zero, which is determined by quantum
geometry, the critical exponent $\gamma$ is introduced to
parameterize the speed of the phase transition.

From the above relation, we can see that when $T=T_{c}$, the
correlation length $R$ is divergent and $T<T_{c}$, $R$ is created,
the transition happens and enters into the usual radiation phase of
standard cosmology.

In holographic cosmology, the holographic phase has a nearly
divergent length, which is required to assure all interesting modes
observed today are in causal contact before transition. This solves
the horizon problem. In the next section, we will mainly discuss the
holographic thermal non-Gaussianity.

\section{The holographic thermal non-Gaussianity}

In this section, our purpose is to discuss the non-Gaussianity of
thermal fluctuations in holographic cosmology. Before discussing the
non-Gaussianity, a key question which has to be addressed is on what
scale the initial conditions should be imposed. We note that in Ref.
\cite{thermalChen}, they hold the thermal horizon $R$ as a parameter
during the calculation. They find that if thermal horizon is smaller
than Horizon scale at the horizon crossing, thermal fluctuations can
lead to a large non-Gaussianity. In Ref. \cite{non-commutative2},
the thermal correlation length $R=T^{-1}$ is adopted, which is a
lower bound and so will lead to larger non-Gaussianity. In our work,
we will adopt the Hubble scale $R=H^{-1}$, beyond which causality
prohibits local causal interactions \cite{causality}. Therefore we
will calculate at $R=H^{-1}$, i.e. $k=a/R=aH$. This means that when
$T=T_{c}$, $R(T)$ is infinite, $H\simeq 0$. Next, we will firstly
calculate the 2-point correlations, 3-point correlations and the
power spectrum. Then we give the non-Gaussianity of holographic
thermal fluctuations.

Fluctuations in a thermal ensemble can be determined from the
thermodynamic partition function
\begin{equation}
  Z=\sum_r e^{-\beta E_r}~,
\end{equation}
where the summation runs over all states, $E_{r}$ is the energy of
the state, and $\beta=T^{-1}$. Let $U$ represents the total energy
inside region $\mathcal{R}$. Then the average energy of the system
is given by
\begin{equation}
  \langle U \rangle = \langle E \rangle = \frac{\sum_{r}E_{r} e^{- \beta E_{r}}}{\sum_{r} e^{- \beta E_{r}}} =  -\frac{d \log Z}{d\beta}~,
\end{equation}

The 2-point correlation function for the energy fluctuations $\delta
U\equiv U-\langle U \rangle$ is given by
\begin{equation}\label{energy2}
  \langle \delta U^{2} \rangle=\frac{d^{2}\log Z}{d\beta^{2}}=-\frac{dU}{d\beta}=T^{2}c_{\langle A\rangle}~,
\end{equation}

Similarly, the 3-point correlation function can be expressed as
\begin{equation}\label{energy3}
  \langle \delta U^{3} \rangle=-\frac{d^{3}\log Z}{d\beta^{3}}=\frac{d^{2}U}{d\beta^{2}}=T^{3}(2c_{\langle A\rangle}+Tc'_{\langle A\rangle})~.
\end{equation}
where prime denotes the derivative with respect to the temperature
$T$.

Since the energy density perturbations $\delta \rho = \delta E/V$.
Using (\ref{heat}) and (\ref{energy2}), the 2-point correlation
functions of $\delta \rho$ is given as
\begin{equation}\label{rho2}
  \langle\delta\rho^2\rangle=\frac{\langle\delta
  U^2\rangle}{V^{2}}=\frac{T^{2}c_{\langle A \rangle}}{R^{6}}=\frac{4\pi bT^{2}}{\hbar
  GR^{4}}~.
\end{equation}

Now we calculate the 3-point correlation function of $\delta \rho$.
We note that $c'_{A}=0$. Using (\ref{heat}) and (\ref{energy3}), the
3-point correlation function of $\delta \rho$ can be expressed as
\begin{equation}\label{rho3}
  \langle\delta\rho^3\rangle=\frac{\langle\delta
  U^3\rangle}{V^{3}}=\frac{T^{3}(2c_{\langle A \rangle}+Tc'_{\langle A \rangle})}{R^{9}}=\frac{8\pi bT^{3}}{\hbar GR^{7}}~.
\end{equation}

Performing the fourier transformation, the density fluctuations
$\delta\rho_{k}$ in momentum space can be related the fluctuation
$\delta\rho$ in position space by
\begin{equation}\label{rhok}
  \delta\rho_{k}=k^{-{3\over2}}\delta\rho~.
\end{equation}

In longitudinal gauge (see \cite
{longitudinal1,longitudinal2,longitudinal3}), and in the absence of
anisotropic matter stress, the metric takes the form
\begin{equation}\label{metric}
  ds^{2}=a^{2}(\eta)[-d\eta^{2}(1-2\Phi)+(1+2\Phi)dx^{2}]~,
\end{equation}
where $\Phi$ represents the fluctuations in the metric. Since during
the phase transition, $R(T)$ is infinite, $H\simeq 0$, we can have
$k\gg H$, which means that the perturbations are deep in the
horizon. Thus the Eq. (\ref{metric}) of metric perturbation may be
reduced to the Poisson equation
\begin{equation}
  k^{2}\Phi_{k}=4 \pi G a^{2}\delta\rho_{k}~.
\end{equation}

Using the horizon crossing condition $k=a/R=aH$, the Poisson
equation can be expressed as
\begin{equation}\label{Phik}
  \Phi_{{k}L} = 4\pi G \delta\rho_{k}R^2~.
\end{equation}

Using (\ref{rho2}), (\ref{rhok}) and (\ref{Phik}) , the 2-point
correlation for $\Phi_{k}$ is given
\begin{equation}\label{Phi2}
  \langle\Phi_{k}^{2}\rangle =\frac{(4\pi)^{3}GbT^{2}}{\hbar}k^{-3}~.
\end{equation}

Similarly, the 3-point function for $\Phi_{k}$ is expressed as
\begin{equation}\label{Phi3}
  \langle\Phi_{k}^{3}\rangle = \frac{2(4\pi)^{4}G^{2}bT^{3}}{\hbar R}k^{-\frac{9}{2}}~.
\end{equation}

If we neglect the variation in $T$ during the transition (fixed
temperature), the power spectrum for $\Phi_{k}$ can be written as
\begin{equation}\label{powerholography1}
  P_{\Phi} \equiv \frac{k^{3}}{2\pi^{2}}\langle\Phi_{k}^{2}\rangle= 32\pi b\frac{T_{c}^{2}}{T_{Pl}^{2}}~.
\end{equation}
where $T_{Pl}=\sqrt{\frac{\hbar}{G}}$ is Planck temperature. Owing
to the energy proportional to the area, that is a scale-invariant
spectrum but not the white noise.

Next, we will calculate the holographic thermal non-Gaussianity.
Note that $\Phi$ perturbs CMB through the so-called Sachs-Wolfe
effect \cite{Sachs-Wolfe}. However, it is useful to introduce a
second variable, $\zeta$, which is the primordial curvature
perturbation on comoving hypersurfaces \cite{Bardeen80,Bardeen83}.
Then the non-Gaussianity estimator $f_{NL}^{equil}$ can be
calculated theoretically by
\begin{equation}\label{fdefine}
  f_{NL}^{equil} = \frac{5}{18}k^{-\frac{3}{2}}\frac{\langle {\zeta}_{k}^{3}\rangle}{\langle{\zeta}_{k}^{2}\rangle \langle{\zeta}_{k}^{2}\rangle}~.
\end{equation}

The variables $\Phi$ and $\zeta$ are related by
\begin{equation}\label{Phizeta1}
  \zeta=\Phi-\frac{H}{\dot{H}}(\dot{\Phi}+H\Phi)~.
\end{equation}

The variable $\zeta$ remaining nearly constant at super-horizon
scales for adiabatic fluctuations but $\Phi$ not
\cite{longitudinal1}. However, if the equation of state is constant,
then $\Phi$ also remains constant at super-horizon. Therefore the
relation (\ref{Phizeta1}) reduces to \cite{Kodama},
\begin{equation}\label{Phizeta2}
  \zeta=\frac{5+3\omega}{3+3\omega}\Phi.
\end{equation}

We also note that the primordial curvature variable $\zeta$ is
independent of $\omega$, but the variable $\Phi$, which perturbs the
CMB, changes as $\omega$ changes. For more detailed discussion on
the variables $\zeta$ and $\Phi$, we can refer to
\cite{longitudinal1,NGReview,non-commutative2}.

Therefore, combining Eqs. (\ref{Phi2}), (\ref{Phi3}),
(\ref{fdefine}) and (\ref{Phizeta2}), the non-Gaussianity estimator
$f_{NL}^{equil}$ can be expressed as
\begin{equation}\label{f1}
  f_{NL}^{equil} = \frac{5(1+\omega)\hbar}{48(5+3\omega) \pi^{2}bT_{c}R}= \frac{5(1+\omega)\hbar H}{48(5+3\omega)
  \pi^{2}bT_{c}}~.
\end{equation}

Using the condition that modes exit the Hubble radius $k=a_{c}H$ and
$k_{0}=a_{0}H_{0}$, and considering the relationship between scale
factor and temperature $\frac{a_{c}}{a_{0}}=\frac{T_{0}}{T_{c}}$ in
radiation-dominated era, we can estimate the non-Gaussianity
estimator $f_{NL}^{equil}$ to be
\begin{equation}\label{f2}
  f_{NL}^{equil} = \frac{5(1+\omega)\hbar}{48(5+3\omega) \pi^{2}b}\frac{H_{0}}{T_{0}}\frac{k}{k_{0}} = \frac{5(1+\omega)\hbar}{48(5+3\omega) \pi^{2}b}\times
  10^{-30}\frac{k}{k_{0}}~,
\end{equation}
where $H_{0}$, $T_{0}$, and $k_{0}$ represent today's values,
$a_{c}$ is the scale factor during the phase transition and we have
neglected the variation in $a_{c}$.

From the above relation, we find that if $\omega\rightarrow -1$, the
non-Gaussianity will be suppressed as in usual inflationary phase.
However, if the matter is phantom-like, the non-Gaussianity
$f_{NL}^{equil}$ can reach  ${\cal O}(1)$ or larger than ${\cal
O}(1)$ by fine tuning of the equation of state $\omega$. Especially
in the limit $\omega\rightarrow -5/3$, the non-Gaussianity can be
very large. Therefore, the thermal holographic non-Gaussianity
depends on what kind of matter in holographic phase (Phase
\textrm{I}), in which these fluctuations propagate to the primordial
curvature fluctuation $\zeta$ or the potential $\Phi$ when the
classical metric emerges in Phase \textrm{II}, so that the large
thermal non-Gaussianity occurs in holographic phase imprints on
Phase \textrm{II}. Also, we must point out that when $\omega < -1$,
the non-Gaussianity estimator $f_{\rm NL}^{equil} <0$. Therefore the
large non-Gaussianity in holographic cosmology will be negative.

Moreover, the non-Gaussianity estimator $f_{NL}^{equil}$ depends
linearly on the mode $k$. This is similar with string gas cosmology
\cite{thermalSGC}, where the non-Gaussianity estimator
$f_{NL}^{equil}$ also depends linearly on the mode $k$, but very
different from what happens in inflationary models, where the
non-Gaussianities are almost scale invariant.  As pointed out in
Ref. \cite{thermalSGC}, depending linearly on the mode $k$ means
that modes reentering the Hubble radius earlier have a larger
non-Gaussianity. Therefore, if non-Gaussianity with an amplitude
growing linearly with $k$ were to be detected, then thermal
holographic cosmology with phantom-like matter should be desirable.

In addition, we must point out that the non-Gaussianity estimator
$f_{NL}^{equil}$ also depends sensitively on the parameter $b$. If
the parameter $b$ is small, then the non-Gaussianity could be large.
However, it will be confronted with the same problem as in string
gas cosmology, in which if the string scale is decreased to the TeV
scale, in order to obtain a power spectrum with reasonable
amplitude, a fine-tuning on the temperature $T$ of the thermal
string gas should be required \cite{thermalSGC}. In this one would
also require a fine-tuning on the phase transition temperature
$T_{c}$.

Now, in fact, if we depict more precisely the phase transition by
the relation (\ref{RandT}), the spectrum in phase \textrm{II} cannot
be exactly scale-invariant and the non-Gaussianity estimator
$f_{NL}^{equil}$ cannot be exactly depends linearly on the mode $k$.
In this case, the spectrum can be expressed as more precisely
\begin{equation}\label{powerholography2}
  P_{\Phi} = 32\pi b\frac{T_{c}^{2}}{T_{Pl}^{2}}[1-(\frac{l_{0}}{R})^{\frac{1}{\gamma}}]^{2}= 32\pi b\frac{T_{c}^{2}}{T_{Pl}^{2}}[1-(l_{0}T_{c}\frac{H_{0}}{T_{0}}\frac{k}{k_{0}})^{\frac{1}{\gamma}}]^{2}~.
\end{equation}

Similarly, the non-Gaussianity estimator $f_{NL}^{equil}$ takes the
form
\begin{equation}\label{fholography2}
  f_{NL}^{equil} = \frac{5(1+\omega)\hbar}{48(5+3\omega)
  \pi^{2}b}\frac{H_{0}}{T_{0}}\frac{k}{k_{0}}[1-(l_{0}T_{c}\frac{H_{0}}{T_{0}}\frac{k}{k_{0}})^{1/\gamma}]^{-1}
  \simeq \frac{5(1+\omega)\hbar}{48(5+3\omega)
  \pi^{2}b}\frac{H_{0}}{T_{0}}\frac{k}{k_{0}}[1+(l_{0}T_{c}\frac{H_{0}}{T_{0}}\frac{k}{k_{0}})^{1/\gamma}].
\end{equation}

We can see that the non-Gaussianity estimator $f_{NL}^{equil}$
depends on $k$ in a more complicated manner, which is very different
from the string gas cosmology.

\section{Conclusion and discussion}

In this paper, we have calculated the non-Gaussianity parameter
$f_{NL}^{equil}$ for thermal fluctuations in holographic cosmology
with the use of the strategy developed in \cite{thermalChen}. Since
the energy is proportional to the area, the thermal holographic
non-Gaussianity with an amplitude growing linearly with $k$ is
obtained if we neglect the variation in $T$ during the transition
(fixed temperature). Furthermore, if we assume the dependence of $R$
on $T$ as (\ref{RandT}) during the transition, then we find that the
non-Gaussianity estimator $f_{NL}^{equil}$ depends on $k$ in a more
complicated manner.

Furthermore, the thermal holographic non-Gaussianity depends on what
kind of matter in Phase \textrm{I}. If $\omega\rightarrow -1$, the
non-Gaussianity will be suppressed as in usual inflationary phase.
However, if the matter is phantom-like, the non-Gaussianity
$f_{NL}^{equil}$ can reach  ${\cal O}(1)$ or larger than ${\cal
O}(1)$ by fine tuning of the equation of state $\omega$. Especially
in the limit $\omega\rightarrow -5/3$, the non-Gaussianity can be
very large. In fact, from the expression (\ref{fdefine}) of the
non-Gaussianity estimator $f_{NL}^{equil}$ and the relation
(\ref{Phizeta2}), we can immediately see that when the
$\omega\rightarrow -5/3$, $\zeta\rightarrow 0$, so the
non-Gaussianity estimator $f_{NL}^{equil}\rightarrow \infty$. If
$\omega\rightarrow -1$, $\zeta\rightarrow \infty$, the
non-Gaussianity estimator $f_{NL}^{equil}\rightarrow 0$. In single
field slow roll inflationary model, in order to obtain the
scale-invariant spectrum, near deSitter spacetime
($\omega\rightarrow -1$) is required, then $\zeta\rightarrow
\infty$, the non-Gaussianity estimator $f_{NL}^{equil}\rightarrow
0$. Therefore, the non-Gaussianity is suppressed in single field
slow roll inflationary model. However, in holographic cosmology, the
scale-invariant spectrum is obtained by the specific heat scaling as
area in holographic phase. Since the holographic phase transition is
determined by geometry rather than the matter content
\cite{thermalholography1}, the large non-Gaussianity can be achieved
by an appropriate choice of $\omega$. We also note that the phase
transition in string gas cosmology depends on the matter content (in
the Hagedorn phase $\omega=0$ and in the radiation dominated
universe $\omega=1/3$ \cite{thermalSGC,Hagedorn}) but not geometry.
So the non-Gaussianity in string gas cosmology is small unless fine
tuning of the string scale \cite{thermalSGC}. Therefore, in
holographic phase, just as the geometry change abruptly, the large
non-Gaussianity can be achieved. However, it is still an open
question on what kind of matter will be dominant in early
non-geometric phase. It will be very interesting to furthermore
explore this question.

It is interesting that the large non-Gaussianity can be achieved by
an appropriate choice of $\omega$ in thermal holographic cosmology.
Next, we will also consider the thermal non-Gaussianity in
semiclassical loop cosmology and Near-Milne universe
\cite{preparation}.

\section*{Acknowledgement}
We are grateful to Robert Brandenberger  and Yi Wang for their
replies with helpful suggestions and comments. This work is partly
supported by NSFC(Nos.10663001,10875057), JiangXi SF(Nos. 0612036,
0612038), Fok Ying Tung Eduaction Foundation(No. 111008) and the
key project of Chinese Ministry of Education(No.208072). We also
acknowledge the support by the Program for Innovative Research
Team of Nanchang University.

\end{document}